\documentclass[a4paper,10pt]{article}
\usepackage[a4paper]{geometry}

\usepackage{amsmath}
\usepackage{amssymb}
\usepackage{xcolor}
\usepackage{comment}
\usepackage[T1]{fontenc}
\usepackage[utf8]{inputenc}
\usepackage[english]{babel}
\usepackage{cite}
\usepackage{soul}
\usepackage{mathtools} 

\usepackage{graphicx}
\usepackage[colorlinks=true,citecolor=blue,urlcolor=blue,linkcolor=blue]{hyperref} 
\usepackage{amsthm}
\newtheorem{conjecture}{Conjecture}

\newtheorem{proposition}{Proposition}

\newcommand{\CHM}[1]{\mathbb{H}(#1)} 
\newcommand{\BH}[2]{{\rm B}\mathbb{H}(#1,#2)} 
\newcommand{\TUCHM}[1]{\mathbb{H}^{\rm 2}(#1)} 
\newcommand{\KUCHM}[2]{\mathbb{H}^{\rm #2}(#1)} 
\newcommand{\DUCHM}[1]{\mathbb{H}^{\rm R}(#1)} 
\newcommand{\GUCHM}[1]{\mathbb{H}^{\rm \Gamma}(#1)} 
\newcommand{\TUBH}[2]{{\rm B}\mathbb{H}^{\rm 2}(#1,#2)} 
\newcommand{\KUBH}[3]{{\rm B}\mathbb{H}^{\rm #3}(#1,#2)} 
\newcommand{\DUBH}[2]{{\rm B}\mathbb{H}^{\rm R}(#1,#2)} 
\newcommand{\GUBH}[2]{{\rm B}\mathbb{H}^{\rm \Gamma}(#1,#2)} 

\title{Multi-Unitary Complex Hadamard Matrices}
\author{Wojciech Bruzda$^{1}$\footnote{w.bruzda@uj.edu.pl}, 
Grzegorz Rajchel-Mieldzio\'{c}$^{2}$, Karol \.{Z}yczkowski$^{1,3}$}
\date{
{\small
$^1$Center for Theoretical Physics, Polish Academy of Sciences,\\ al. Lotnik\'{o}w 32/46, 02-668 Warszawa, Poland
\\
\smallskip
$^2$NASK National Research Institute, ul. Kolska 12, 01-045 Warszawa, Poland
\\
\smallskip
$^3$Faculty of Physics, Astronomy and Applied  Computer Science, Jagiellonian University,\\ ul. {\L}ojasiewicza 11, 30-348, Krak\'{o}w, Poland
}\\
\medskip
{\small 2024/06/14}}

\begin{document}
\maketitle

\begin{abstract}
We analyze the set of real and complex Hadamard matrices with additional symmetry constrains.
In particular, we link the problem of existence of {\sl maximally entangled multipartite states} of $2k$ subsystems with $d$ levels each to the set of complex Hadamard matrices of order $N=d^k$.
To this end, we investigate possible subsets of such matrices
which are, {\sl dual}, {\sl strongly dual} ($H=H^{\rm R}$ or $H=H^{\rm\Gamma}$),  
{\sl two-unitary} ($H^R$ and $H^{\Gamma}$ are unitary), 
or {\sl $k$-unitary}.
Here $X^{\rm R}$ denotes reshuffling of a matrix $X$ describing a bipartite system,
and $X^{\rm \Gamma}$ its partial transpose.
Such matrices find several applications in quantum many-body theory, tensor networks and classification of multipartite quantum entanglement
and imply a broad class of analytically solvable quantum models in $1+1$ 
dimensions.
\end{abstract}


\section{Introduction}

\subsection{Complex Hadamard Matrices}

A square matrix $H$ with $\pm 1$ entries is said to be a {\sl (real) Hadamard matrix}~\cite{Ha93}, if its rows (or columns) are mutually orthogonal.
This definition can be generalized by expanding the entries of $H$ to a unit circle on the complex plane. Such matrices are called {\sl complex Hadamard}
matrices and they are the main subject of this paper. 
We write
\begin{equation}
H\in\CHM{N} \ \Longleftrightarrow \ \forall\,j,k : |H_{jk}| = 1 \ \text{and} \ HH^{\dagger}=N\,\mathbb{I},
\end{equation}
where $\CHM{N}$ denotes the set of complex Hadamard matrices (CHM).
These matrices, unitary up to rescaling, are extensively used in contemporary theoretical physics and mathematics~\cite{We01, Iv81, WF89, Ba22}.

Every $H\in\CHM{N}$ is invariant with respect to multiplication by monomial unitary matrices $M=PD$,
where $P$ is a permutation matrix and $D$ is a unitary diagonal matrix~\cite{Ha97}.
In other words, both matrices $H$ and $H'=MHM'$ belong to the same orbit and are called {\sl equivalent}, written $H\simeq H'$.
The classification of all distinct orbits of CHM for a fixed dimension $N$, completed by Haagerup in cases $N=2,3,4,5$~\cite{Ha97}, remains open for $N\geqslant 6$~\cite{TZ06, MS08, Br23}.
Sometimes it is convenient to express a matrix $H$ in the {\sl dephased} ({\sl normalized}) form, in which its first row and first column consist of ones.
The remaining submatrix of order $(N-1)^2$ is called the {\sl core} of $H$.

In the class of CHM of size $N$ one distinguishes a proper subset $\BH{N}{q}\subset\CHM{N}$,
called Butson matrices~\cite{Bu62,Bu63}.
Given integer $q>1$, we write that $B\in\BH{N}{q}$, if
all entries of $B$ are $q^{\rm th}$ roots of unity, i.e.\ $B_{jk}^q=1$
for any $j$ and $k$.
A comprehensive study of Butson-type matrices and their monomial equivalence classification in low dimensions has been performed in Ref.~\cite{LOS19}.
An Appendix to that work, called \texttt{Butson Home}, is presented as the online catalog~\cite{BHAalto} and
contains precalculated Butson matrices provided in the logarithmic form.
In the following sections, these prefabricated units will serve as an input to the procedures
exploring the sets of dual or multi-unitary Hadamard matrices.
Throughout the paper, symbol $N$ shall denote the square dimension $N=d^2$ for $d\geqslant 2$, unless stated otherwise, and
we mainly focus on two cases for $d=3$ and $4$.

\subsection{Subsets of Complex Hadamard Matrices}
\label{sec:CHM_subsets}

We introduce several proper subsets of $\CHM{N}$, which will be subjected to investigation in the following sections. 

Any Hadamard matrix $H$ of dize $N=d^2$ can be written as a tensor $H_{jk}=H_{ab;cd}$.
We define the operations of {\sl reshuffling} and {\sl partial transpose} as $H^{\rm R}_{ab;cd}=H_{ac;bd}$ and $H^{\rm\Gamma}_{ab;cd}=H_{ad;cb}$, respectively. 
Note that both operations, in general, do not preserve unitarity.

We say that $U\in\mathbb{U}(N)$ is {\sl dual unitary} matrix if both $U$ and (independently) $U^{\rm R}$ are unitary. 
Similarly, $U$ is called {\sl $\Gamma$-dual unitary} if both $U$ and
$U^{\rm\Gamma}$ are unitary.\footnote{Sometimes we interchangeably use the notion of {\sl dual unitarity} and {\sl ${\rm R}$-dual unitarity}, but {\sl $\rm\Gamma$-duality} shall always be written explicitly.}
When restricted to the set of (complex) Hadamard matrices, we define two subsets
$\DUCHM{N}$ and $\GUCHM{N}$ containing aforementioned matrices.
Special cases where $U=U^{\rm R}$ or $U=U^{\rm\Gamma}$ ({\sl strong duality}) will be considered in  
Section~\ref{sec:dual_self_dual_CHM}.
Furthermore, if matrix $U$ remains unitary after both operations;
reshuffling $U^{\rm R}\in\mathbb{U}(N)$ and partial transpose $U^{\rm\Gamma}\in\mathbb{U}(N)$, we call it a {\sl $2$-unitary} matrix (sometimes written {\sl two-unitary}).
Such matrices in the set $\CHM{N}$ are elements of the subset denoted by $\TUCHM{N}$.
A natural extension of $2$-unitarity is
$k$-unitarity defined for a 
system consisting of $2k$ parties,
each with $d$ internal levels ($d$ being the dimension of a local Hilbert space), see Section~\ref{sec:mu_AME} and Ref.~\cite{GALRZ15}.
Hence, a $k$-unitary matrix of order $d^k$ preserves unitarity regardless of which rearrangement
of multi-index of $U$ has been applied. 
Hadamard matrices which are $k$-unitary will belong to the set denoted by $\KUCHM{N}{k}$ and we will pay special attention to $k=2$.

The following inclusion relations hold:
\begin{equation}
\KUCHM{N}{k} \subset \DUCHM{N} \subset \CHM{N} \subset \sqrt{N}\mathbb{U}(N).
\end{equation}
For any $q>2$, a similar structure inside the Butson class can be written,
\begin{equation}
\KUBH{N}{q}{k} \subset \DUBH{N}{q} \subset \BH{N}{q} \subset \CHM{N} \subset \sqrt{N}\mathbb{U}(N).
\end{equation}
In both relations the sets $\DUCHM{N}$ and $\DUBH{N}{q}$ can
be replaced by $\GUCHM{N}$ and $\GUBH{N}{q}$, respectively.
As it will be shown, none of the sets $\DUCHM{N}$, $\GUCHM{N}$ and $\KUCHM{N}{k}$ is empty,
and it is rather easy to obtain such matrices
in dimension $N=9$ and $N=16$, both numerically and analytically.
Numerical methods used in this paper are described in
Appendix~\ref{app:numerical_methods}.

A straightforward criterion to confirm duality and $2$-unitarity of a matrix is checking that appropriate matrices remain unitary after rearrangements of their entries. Instead, one can simply calculate the associated linear entropy of any matrix $X$ of size $d^2$,
defined as
\begin{equation}
S(X)=\frac{d^2}{d^2-1}\left(1-\frac{{\rm Tr}(XX^{\dagger}XX^{\dagger})}{{\rm Tr}^2(XX^{\dagger})}\right)\in[0,1].\label{SL_entropy}
\end{equation}
By construction, for any unitary $U\in\mathbb{U}(d^2)$ one has $S(U)=1$.

For brevity, we are going to write the triplet $(S(U),S(U^{\rm R}),S(U^{\rm\Gamma}))$ as $S(U)=(a,b,c),$
where $a$, $b$, $c$ correspond to the entropies of $U$, $U^{\rm R}$
and $U^{\rm\Gamma}$, respectively.
Hence, $S(U)=(1,1,1)$ shall denote a $2$-unitary matrix, while
$S(U)=(1,1,c)$ and $S(1,b,1)$ with $b,c\in[0,1)$ stand
for strictly dual ($\rm R$-dual) and $\rm\Gamma$-dual matrices $U$.

\subsection{\texorpdfstring{Two-unitary Matrices of order $d^2$ and Local Unitary Invariants}{}}\label{section:LUI}

In the set of $2$-unitary matrices defined in Sec.~\ref{sec:CHM_subsets} (not necessarily in the Hadamard subset)
we distinguish special operations that preserve the property of $2$-unitarity.
Similarly to the idea of Hadamard orbits, we say that two $2$-unitary matrices $X$ and $X'$ of order $d^2$
are {\sl locally unitarily equivalent}, if there exist
four unitary matrices $U_1$, $U_2$, $U_3$ and $U_4$ in $\mathbb{U}(d)$, such that
$(U_1\otimes U_2)X(U_3\otimes U_4)=X'$, written $X\simeq_{\rm LU} X'$.
This means that the matrix $X\in\mathcal{U}(d^2)$ describing a four-partite system in Hilbert space $\mathcal{H}_{d^2}$, see Eq.~(\ref{eq:tensor_expansion}), composed of
$d$-dimensional subsystems, does not change qualitatively if subjected to local rotations in four subspaces $\mathcal{H}_d$. 
From the operational perspective, local rotations are implemented by means of local unitary operations in individual laboratories.
Note that LU-equivalence differs from the standard Hadamard equivalence~\cite{Ha97} and, in general, Hadamard monomial operations strongly affect the properties of $2$-unitarity. 

Given $d\geqslant 2$, a natural question appears,
how many LU-equivalence classes there exist in the set of $2$-unitary matrices. The problem is investigated in Ref.~\cite{RRKL23}. In order to illustrate the idea,
consider the following permutation matrix of order nine,
\begin{equation}
P_9=[1,9,5,6,2,7,8,4,3],\label{AME43P9}
\end{equation}
where each number represents the position of unity in consecutive columns of $P_{9}$.
One can easily confirm that for any four three-dimensional unitary matrices $U_j\in\mathbb{U}(3)$,
matrix $(U_1\otimes U_2)P_9(U_3\otimes U_4)$ is $2$-unitary. 
In other words, the linear entropy
achieves its maximal values for every realignment of $P_9$, i.e. $S(P_9)=(1,1,1)$.
In this case it can be proven~\cite{RRKL23}, that there exists only one LU-equivalence class,
so that the matrix $P_9$ is its simplest representative.

In dimension $d=4$, there are at least two LU-equivalence classes in the set of $2$-unitary matrices.
The first one is generated by the following permutation matrix~\cite{RAL22},
\begin{equation}
P_{16}=[1,16,6,11,15,2,12,5,8,9,3,14,10,7,13,4],
\label{P16_matrix}
\end{equation}
where each number was encoded like in Eq.~\eqref{AME43P9}, corresponding to the case of $P_9$.
The explicit form of a representative of the other class, which is orthogonal up to rescaling, reads
\begin{equation}
O_{16}=
{\tiny\left[\begin{array}{cccc|cccc|cccc|cccc}
+ & . & . & . &  . & + & . & . &  . & . & -& . &  . & . & . &-\\
. & + & . & . &  -& . & . & . &  . & . & . & -&  . & . &- & .\\
. & . &- & . &  . & . & . & + &  -& . & . & . &  . &- & . & .\\
. & . & . &- &  . & . & + & . &  . & + & . & . &  + & . & . & .\\
\hline
. & + & . & . &  -& . & . & . &  . & . & . & + &  . & . & + & .\\
-& . & . & . &  . & + & . & . &  . & . & + & . &  . & . & . &-\\
. & . & . &- &  . & . & + & . &  . &- & . & . &  -& . & . & .\\
. & . &- & . &  . & . & . &- &  + & . & . & . &  . &- & . & .\\
\hline
. & . & -& . &  . & . & . &- &  -& . & . & . &  . & + & . & .\\
. & . & . & + &  . & . & + & . &  . &- & . & . &  + & . & . & .\\
-& . & . & . &  . &- & . & . &  . & . &- & . &  . & . & . &-\\
. &- & . & . &  -& . & . & . &  . & . & . &- &  . & . & + & .\\
\hline
. & . & . &- &  . & . & -& . &  . & -& . & . &  + & . & . & .\\
. & . & + & . &  . & . & . & -&  -& . & . & . &  . &- & . & .\\
. & + & . & . &  + & . & . & . &  . & . & . & -&  . & . & + & .\\
+ & . & . & . &  . &- & . & . &  . & . & + & . &  . & . & . &-\\
\end{array}\right]}.\label{O16_matrix}
\end{equation}

Since there exists only a single LU-equivalence class in the set of $9$-dimensional $2$-unitary matrices,
it implies that all complex Hadamard matrices with this property are connected via local unitary operations by means of four $U_j\in\mathbb{U}(3)$ for $j=1,2,3,4$ -- see Sec.~\ref{sec:2uCHM}.
This is no longer the case for $d=4$ and for other higher dimensions, for which one can distinguish different LU-orbits~\cite{RAL22}.

In the following sections we go beyond permutations and other simple matrices like~\eqref{O16_matrix} 
into the realm of CHM in the search for multi-unitary representatives.


\section{\texorpdfstring{Dual and Self-Dual Hadamard matrices of order $d^2$}{}}
\label{sec:dual_self_dual_CHM}

Unitary operator $U$, for which the reshuffled matrix $U^{\rm R}$ is also unitary is called {\sl dual unitary}.
These operators
gained broad interest due to their facility of modeling dynamics and exact solving nonintegrable many-body systems~\cite{AWGG16, BKP19, PBCP20, BKP20, CL21, RAL20}.
The concept of duality becomes useful for applications in the theory of
quantum cellular automata in $1+1$ and $2+1$ dimensions~\cite{MPW24}.
Dual unitary matrices play a crucial role
in the search of absolutely maximally entangled states~\cite{RBBMLZ22}, as the hypothetical solution
in the form of a $2$-unitary matrix lies on the intersection of two ``dual'' subspaces.

The duality is usually defined in the literature by the operation of reshuffling.
Another, somehow symmetric definition uses transformation defined by partial transposition with respect to $1^{\rm st}$ or $2^{\rm nd}$ subsystem, ${\rm T_1}$ or $\rm T_2$, respectively~\cite{Ne18}. Since, given $X$, one has $(X^{\rm T_1})^{\rm T_2} = (X^{\rm T_2})^{\rm T_1} = X^{\rm T}$, we
restrict our considerations only to $\rm\Gamma \equiv T_2$.

Concrete representations of dual unitaries can be immediately constructed.
Canonical examples include: identity operator, \texttt{CNot} gate, or the permutation matrix $P_9$, determined by Eq.~\eqref{AME43P9}. One can also easily confirm that several Butson matrices~\cite{BHAalto} share the property of $\rm R$- or $\rm\Gamma$-duality.

Let us focus on more complicated objects.
We recall the quantized ``cat map''~\cite{HB80}, in a form borrowed from Ref.~\cite{ASL19},
\begin{equation}
\big(G_{N}(a,b,c)\big)_{jk}\equiv\exp\Big\{\frac{i\pi}{N}\left(aj^2+bk^2+cjk\right)\Big\},
\label{cat-map}
\end{equation}
for $(a,b,c)\in\mathbb{R}^3$ and $1\leqslant j,k\leqslant N=d^2$ with $d\geqslant 3$.
Several properties of such a map is straightforwardly related to the notion of ${\rm R}$- and ${\rm\Gamma}$-duality.
For example matrix $G_9(1,4,-4)$ is ${\rm\Gamma}$-dual
while $G_{16}(1,2,-2)$ is ${\rm R}$-dual one.
Also, two-unitary matrices (see next section) can be easily recovered from this map, for example
$G_9(-4,  -8, -4) \in \TUBH{9}{9}$ or
$G_{49}(1,-1,2) \in\TUBH{49}{98}$.

In order to complete the picture, we shall mention the strong duality with respect to both operations, separately ${\rm R}$ or ${\rm \Gamma}$. 
We call a matrix $X$ of order $N=d^2$ {\sl strongly dual} (or {\sl self-dual}),
if $X=X^{\rm R}$. Similar definition pertains to {\sl strong $\Gamma$-duality} (or {\sl self $\Gamma$-duality}),
if $X=X^{\rm \Gamma}$. Both these properties, however, are volatile and they are not preserved
along LU-equivalence orbits. Suppose $X=X^{\rm R}$ or $X=X^{\rm \Gamma}$, then in general
\begin{equation}
Y=(U_1\otimes U_2)X(U_3\otimes U_4) \quad \Longrightarrow \quad Y\neq Y^{\rm R} \ \text{or} \ Y\neq Y^{\rm \Gamma}
\end{equation}
for $U_j\in\mathbb{U}(d)$. They are not preserved along Hadamard equivalence orbits either.
This makes such matrices very specific points in set $\CHM{N}$.
A currently open problem is, whether it is possible to define special (restricted) orbits for self-dual matrices.

It is not difficult to obtain self-R-dual and self-$\Gamma$-dual matrices both numerically and analytically. 
One particular symmetric example fished out from the collection~\cite{BHAalto}, reads\footnote{Every $\exp\{\cdot\}$ function defining matrices should be understood as acting elementwise.}
\begin{equation}
B=\exp\Bigg\{\frac{2i\pi}{3}
{\scriptsize\left[\begin{array}{ccc|ccc|ccc}
        . & . & . & . & . & . & . & . & .\\
        . & . & . & 1 & 1 & 1 & 2 & 2 & 2\\
        . & . & . & 2 & 2 & 2 & 1 & 1 & 1\\
        \hline
        . & 1 & 2 & . & 1 & 2 & . & 1 & 2\\
        . & 1 & 2 & 1 & 2 & . & 2 & . & 1\\
        . & 1 & 2 & 2 & . & 1 & 1 & 2 & .\\
        \hline
        . & 2 & 1 & . & 2 & 1 & . & 2 & 1\\
        . & 2 & 1 & 1 & . & 2 & 2 & 1 & .\\
        . & 2 & 1 & 2 & 1 & . & 1 & . & 2\\
\end{array}\right]
}
\Bigg\}=B^{\rm R}\in\DUBH{9}{3}.\label{dephased_example}
\end{equation}
The matrix $B$ is self-R-dual with $S(B)=(1,1,0)$ 
as consecutive rows fold to the appropriate $3\times 3$ blocks.
Less trivial constructions and multi-parametric self-dual families (with respect to $\rm R$ and $\rm\Gamma$ operations)
are listed in Appendix~\ref{app:self_dual}.


\section{Two-Unitary Complex Hadamard Matrices}
\label{sec:2uCHM}

Similarly to the case of dual and self-dual CHM, the subset of $2$-unitary matrices offers structures with various properties.
Many such matrices remain unpublished, so for further examples, the interested Reader can consult the GitHub repository~\cite{GitHub}.

Several Hadamard matrices in the literature and the Catalog~\cite{CHMC} are written in the dephased (normalized)
form, where the first row and the first column consist of ones,
and we are going to follow this convention in this paper.
However, this additional technical constraint immediately spoils the property
of multi-unitarity which strongly depends on internal realignments of matrix entries.
An exceptional dephased example has been presented in Eq.~\eqref{dephased_example}.
Nevertheless, when searching for a particular matrix one can start from a matrix in any form and try to
find two-unitary diagonal matrices
$D_L$ and $D_R$ such that
$Y=D_LX_{d^2}D_R\in\TUCHM{d^2}$, where $X_{d^2}$ represents a particular matrix from
$\CHM{d^2}$ for $d=3$ or $d=4$. 

\subsection{\texorpdfstring{Two-Unitary Butson Matrices of Order $N=9$ and $N=16$}{}}

We start the investigation with Butson-type matrices $\BH{N}{q}$, 
and we use the data published in the online catalog~\cite{BHAalto}.
For $N=9$ it is available for $q\in\{3,6,9,10,12,15\}$.
In each case, except for
$\BH{9}{10}$, one can quickly find representatives which, after
transformation by appropriate diagonal matrices, become members of $\TUCHM{9}$.
Let us list just three matrices that are not $2$-unitary, but they can gain this property by unitary rotations represented by particular diagonal unitary matrices.

The matrix $B$ presented in Eq.~\eqref{dephased_example},
for which $S(B)=(1,1,0)$, can be turned into a $2$-unitary with help of the following
diagonal matrix,
\begin{equation}
D=\exp\Big\{{\rm diag}\Big[\frac{2i\pi}{3}\left(0, 0, 0, 0, 4, 2, 0, 2, 4\right)\Big]\Big\},
\end{equation}
so that we have
\begin{equation}
C=D B D^{\dagger}=
\exp \Bigg\{\frac{2i\pi}{3}{\scriptsize\left[\begin{array}{ccc|ccc|ccc}
        . & . & . & . & 2 & 1 & . & 1 & 2\\
        . & . & . & 1 & . & 2 & 2 & . & 1\\
        . & . & . & 2 & 1 & . & 1 & 2 & .\\
        \hline
        . & 1 & 2 & . & . & . & . & 2 & 1\\
        1 & 2 & . & 2 & 2 & 2 & . & 2 & 1\\
        2 & . & 1 & 1 & 1 & 1 & . & 2 & 1\\
        \hline
        . & 2 & 1 & . & 1 & 2 & . & . & .\\
        2 & 1 & . & . & 1 & 2 & 1 & 1 & 1\\
        1 & . & 2 & . & 1 & 2 & 2 & 2 & 2\\
\end{array}\right]}\Bigg\}
\in\TUBH{9}{3}.
\end{equation}
Matrix $C$ is unitary and after dephasing it reveals the original Butson matrix $B$.
By direct inspection, one can prove that both 
$C^{\rm R}$ and $C^{\rm \Gamma}$ are permutation equivalent to $B$.
This implies that $C$ is a $2$-unitary Hadamard matrix of order $9$ with $S(C)=(1,1,1)$.

The collection of Butson matrices for $N=16$ contains
exactly $5$ and $1786763$ records corresponding to cardinalities of the sets $\BH{16}{2}$ and $\BH{16}{4}$, respectively.
Let the symbol $B_{[k]}$ denote the matrix encoded on the $k^{\rm th}$ place (offset) in $\BH{16}{q}$ in~\cite{BHAalto}.
We consider just two initial cases from the latter (bigger) class, which also includes the smaller one.
First observation is rather trivial -- the following product of $B_{[1]}$ and the permutation matrix $P_{16}$ is $2$-unitary,
\begin{equation}
B_{[1]}P_{16}=\exp\Bigg\{i\pi{\scriptsize\left[\begin{array}{cccc|cccc|cccc|cccc}
        .&.&.&.&.&.&.&.&.&.&.&.&.&.&.&.\\
        .&1&.&1&1&.&1&.&.&1&.&1&1&.&1&.\\
        .&1&1&.&1&.&.&1&1&.&.&1&.&1&1&.\\
        .&.&1&1&.&.&1&1&1&1&.&.&1&1&.&.\\
        \hline
        .&1&.&1&1&.&1&.&1&.&1&.&.&1&.&1\\
        .&.&.&.&.&.&.&.&1&1&1&1&1&1&1&1\\
        .&.&1&1&.&.&1&1&.&.&1&1&.&.&1&1\\
        .&1&1&.&1&.&.&1&.&1&1&.&1&.&.&1\\
        \hline
        .&1&1&.&.&1&1&.&1&.&.&1&1&.&.&1\\
        .&.&1&1&1&1&.&.&1&1&.&.&.&.&1&1\\
        .&.&.&.&1&1&1&1&.&.&.&.&1&1&1&1\\
        .&1&.&1&.&1&.&1&.&1&.&1&.&1&.&1\\
        \hline
        .&.&1&1&1&1&.&.&.&.&1&1&1&1&.&.\\
        .&1&1&.&.&1&1&.&.&1&1&.&.&1&1&.\\
        .&1&.&1&.&1&.&1&1&.&1&.&1&.&1&.\\
        .&.&.&.&1&1&1&1&1&1&1&1&.&.&.&.\\
\end{array}\right]}\Bigg\}\in\TUBH{16}{2},
\label{B16_2}
\end{equation}
without introducing any additional modifications or diagonal factors, where $P_{16}$
is defined in Eq.~\eqref{P16_matrix}.

Lastly, consider the $8^{\rm th}$ matrix from the catalog~\cite{BHAalto} of $\BH{16}{4}$,
\begin{equation}
B_{[8]}=\exp\Bigg\{\frac{2i\pi}{4}{\scriptsize\left[\begin{array}{cccc|cccc|cccc|cccc}
. & . & . & . & . & . & . & . & . & . & . & . & . & . & . & .\\
. & . & . & . & . & . & . & . & 2 & 2 & 2 & 2 & 2 & 2 & 2 & 2\\
. & . & . & . & 2 & 2 & 2 & 2 & . & . & . & . & 2 & 2 & 2 & 2\\
. & . & . & . & 2 & 2 & 2 & 2 & 2 & 2 & 2 & 2 & . & . & . & .\\
\hline
. & . & 2 & 2 & . & . & 2 & 2 & . & . & 2 & 2 & . & . & 2 & 2\\
. & . & 2 & 2 & . & . & 2 & 2 & 2 & 2 & . & . & 2 & 2 & . & .\\
. & . & 2 & 2 & 2 & 2 & . & . & . & . & 2 & 2 & 2 & 2 & . & .\\
. & . & 2 & 2 & 2 & 2 & . & . & 2 & 2 & . & . & . & . & 2 & 2\\
\hline
. & 2 & . & 2 & . & 2 & . & 2 & . & 2 & . & 2 & . & 2 & . & 2\\
. & 2 & . & 2 & . & 2 & . & 2 & 2 & . & 2 & . & 2 & . & 2 & .\\
. & 2 & . & 2 & 2 & . & 2 & . & . & 2 & . & 2 & 2 & . & 2 & .\\
. & 2 & . & 2 & 2 & . & 2 & . & 2 & . & 2 & . & . & 2 & . & 2\\
\hline
. & 2 & 2 & . & . & 2 & 2 & . & 1 & 3 & 3 & 1 & 1 & 3 & 3 & 1\\
. & 2 & 2 & . & . & 2 & 2 & . & 3 & 1 & 1 & 3 & 3 & 1 & 1 & 3\\
. & 2 & 2 & . & 2 & . & . & 2 & 1 & 3 & 3 & 1 & 3 & 1 & 1 & 3\\
. & 2 & 2 & . & 2 & . & . & 2 & 3 & 1 & 1 & 3 & 1 & 3 & 3 & 1\\
\end{array}\right]}\Bigg\}.
\end{equation}
This time we find two diagonal unitary
matrices,
\begin{align}
D_L(\alpha_1) &= {\rm diag}\Big\{1,1,1,1,1,1,e^{2i\pi \alpha_1},e^{2i\pi \alpha_1},\omega,\omega,-1,-1,-1,1,-1,1\Big\}\label{DLexample},\\
D_R(\alpha_2) &= {\rm diag}\Big\{1,1,1,1,1,i,1,i,e^{2i\pi \alpha_2},e^{2i\pi \alpha_2},e^{2i\pi \alpha_2},e^{2i\pi \alpha_2},1,i,1,i\Big\}\label{DRexample},
\end{align}
each admitting additional dependence on a single parameter $\alpha_j\in[0,1)$, with $\omega=\exp\{2 i\pi/3\}$, such that
$Y_{16}^{(2)}(\alpha_1,\alpha_2) = D_L(\alpha_1)B_{[8]}P_{16}D_R(\alpha_2)\in\TUCHM{16}$.
Due to the fact that the final matrix depends on two parameters, it does not necessarily belong to the Butson class. In other words,
$Y_{16}^{(2)}(\alpha_1,\alpha_2)\in\BH{16}{4}$ only for particular values of $\alpha_1$ and $\alpha_2$.
Diagonal matrices~\eqref{DLexample} and~\eqref{DRexample} are
not the only possible solutions and it is possible to find entirely different pairs of $(D_L, D_R)$ that bring $B_{[8]}P_{16}$ to $\TUCHM{16}$.
However, in this paper we are not going to solve the problem of the full classification and uniqueness of the solutions.
A similar remark concerns all diagonal matrices presented throughout the paper.

Note that the presence of $P_{16}$ in both cases is crucial.
The set $\BH{16}{4}$ was probed at random, but no matrix $B_{[k]}$,
for hundreds of different values of $1\leqslant k\leqslant 1786763$, could be brought to a $2$-unitary
one with the help of two diagonal matrices and without additional multiplication by $P_{16}$, as in the above examples. 

Another approach to two-unitary Butson-type matrices
was earlier provided in Ref.~\cite{GKL15}, where the authors describe
special graph states which represent AME$(4,d)$ states corresponding
to $\BH{d^2}{d}$ for odd values of $d$.


\subsection{\texorpdfstring{Analytical Results Inferred from CHM Beyond the ${\rm B}\mathbb{H}$ Class}{}}
We recall the matrix of order $9$ constructed by Karlsson~\cite{K16},
\begin{equation}
K_9^{(2)}(\zeta)={\rm circ}\left[\begin{array}{ccc|ccc|ccc}
1 &  x &  x &  y &  u &  w &  y &  w &  u\\
x &  1 &  x &  w &  y &  u &  u &  y &  w\\
x &  x &  1 &  u &  w &  y &  w &  u &  y\\
\end{array}\right],\label{K9_matrix}
\end{equation}
which is symmetric and block circulant, with circulant blocks (BCCB).
This means that the full form of $K_9^{(2)}$ is obtained by a right-circulant shift of three blocks in~\eqref{K9_matrix}.
It admits two-parametric non-affine family depending on two conjugated pairs of real parameters encoded in a single complex parameter $\zeta$,
which can be concisely written as
\begin{equation}
\left.\begin{array}{cc}x\\y\end{array}\right\} = \frac{1}{4}(1+\zeta)\left(1\pm i\sqrt{\frac{16}{|1+\zeta|^2}-1}\right),
\qquad
\left.\begin{array}{cc}u\\w\end{array}\right\} = \frac{1}{4}(1-\zeta)\left(1\pm i\sqrt{\frac{16}{|1-\zeta|^2}-1}\right),
\end{equation}
for 
$\zeta\in\mathcal{D}=\big\{z\in\mathbb{C} : |1-z|\leqslant 4\big\}\cap\big\{z\in\mathbb{C} : |1+z| \leqslant 4\big\}\setminus\{\mp 1\}$. 
Our first observation is that for any value of $\zeta\in\mathcal{D}$,
one obtains a $2$-unitary complex Hadamard matrix, $Y=K_9^{(2)}(\zeta)P_9\in\TUCHM{9}$, where $P_9$ is defined in Eq.~\eqref{AME43P9}.

Another example involves two diagonal matrices
\begin{align}
D_L&=\exp\Big\{{\rm diag}\Big[\frac{2 \pi i}{3}\left(0,1,1,1,1,0,0,2,0\right)\Big]\Big\}\quad\text{and}\\
D_R&=\exp\Big\{{\rm diag}\Big[\frac{2 \pi i}{3}\left(0,1,1,1,0,1,2,2,1\right)\Big]\Big\},
\end{align}
which, upon application to the tensor product $F_3\otimes F_3$, yield a 2-unitary matrix, $D_L(F_3\otimes F_3)D_R\in\TUCHM{9}$. 
These two matrices allow us to construct a 4-parameter family of 2-unitary matrices, $D_LF_9^{(4)}(\alpha_1,\alpha_2,\alpha_3,\alpha_4)D_R\in\TUCHM{9}$, with arbitrary values of affine parameters $\alpha_j$, $j\in\{1,2,3,4\}$.
In both cases, the proof is straightforward and exploits the observation that all matrices used in the definition of $2$-unitarity are indeed complex Hadamard ones or, equivalently, the vector of linear entropies is (maximally) flat.

In contrast, so far, no similar examples were found for other known matrices: $B_9^{(0)}$, $N_9^{(0)}$,
$S_9^{(0)}$ nor $Y_9^{(0)}$~\cite{BN06, NW12, Br23}.
All these matrices are characterized by vanishing defect and do not admit any internal parameterization.
This suggests the following statement.
\begin{conjecture}
Only non-isolated complex Hadamard matrices of order $N=d^2$ provide two-unitary structures.
\label{isolated-conjecture}
\end{conjecture}
It can be intuitively understood since $2$-unitarity requires a sort of flexibility in the matrix.
Isolated structures have rigidly fixed elements and might not provide enough degrees of freedom
to allow for additional internal rearrangements.
Moreover, in the class $\BH{16}{4}$ discussed by Lampio et al.~\cite{BHAalto}, there are exactly
$7978$ matrices with vanishing value of defect (and very likely there are many others which are implicitly isolated,
i.e. they are isolated despite the fact their defect is non-zero).
A large number of such matrices was picked at random from the set,
and none of them reflected properties of $2$-unitarity when assisted with two diagonal matrices and two different permutations.
These numerical observations encourage us to advance Conjecture~\ref{isolated-conjecture}.

Investigation of all known $16$-dimensional matrices from the Catalog~\cite{CHMC}, namely
\begin{equation}
\Big\{H_{16A}\simeq\bigotimes_{j=1}^4 H_2, \ H_{16B,C,D,E}, \ F_4\otimes F_4,\ S_8\otimes H_2, \ F_8\otimes H_2, \ S_{16}^{(11)}, \ F_{16}^{(17)}\Big\},
\end{equation}
did not reveal any $2$-unitary elements belonging to $\TUCHM{16}$.
However, this does not imply that they do not exist.
It might still happen that a particular choice of permutation matrices, $X\to P_L X P_R$, shall bring one of these matrices to the desired form.

Finally, for $N=16$, the modified Sinkhorn algorithm (supplied with a random-Gaussian seed) finds hundreds of different examples that belong to $\TUCHM{16}$. 
At this stage it is hard to distinguish and classify them.
Some of these examples form families that can be written analytically. One additional family is presented
in Appendix~\ref{app:2UCHM_16}.


\subsection{\texorpdfstring{Local unitary construction of $2$-Unitary Hadamard matrices}{}}

Let us recall once again the permutation matrix
$P_9$ of order $9$, defined in Eq.~\eqref{AME43P9}. Local unitary equivalence relation (LU-equivalence) allows us to act on $P_9$ locally, obtaining
\begin{equation}
Q=(U_1\otimes U_2) P_9 (U_3\otimes U_4),
\end{equation}
where four $U_j\in\mathbb{U}(3)$ are unitary matrices of order three. Resulting matrix $Q\in\mathbb{U}(9)$
preserves the property of being $2$-unitary, but does not need to belong to $\CHM{9}$, for instance,
\begin{equation}
(F_3\otimes F_3)P_9(F_3\otimes F_3)\not\in\CHM{9}.
\end{equation}
Only some special choices of $U_j$ provide $2$-unitary Hadamard matrices. The most general form reads
\begin{equation}
\big(M_{a}F_3^{m_1} M_{b}\otimes M_{c}F_3^{m_2} M_{d}\big)P_9\big(M_{e}F_3^{m_3} M_{f}\otimes M_{g}F_3^{m_4} M_{h}\big)\in\TUCHM{9},\label{LU-2UCHM9}
\end{equation}
where $m_j\in\{0,1\}$, $m_1+m_2+m_3+m_4=2$,
and $M_{x}$ for $x\in\{a,b,...,h\}$ are eight general monomial matrices of the form 
$M_{x}=D_NP_N$ with arbitrarily chosen unitary diagonal $(D_N)$ and permutation $(P_N)$ matrices of order $N$. 
In other words, exactly two of the positions denoted by ellipsis ``$...$'' in
\begin{equation}
\big(M_{a}... M_{b}\otimes M_{c}... M_{d}\big)P_9\big(M_{e}... M_{f}\otimes M_{g}... M_{h}\big)
\end{equation}
should be filled by the Fourier matrix $F_3$, while the other two by $\mathbb{I}_3$, leaving $\binom{4}{2}=6$ such possibilities in total.

Similarly, one can write
\begin{equation}
\big(M_{a}... M_{b}\otimes M_{c}... M_{d}\big)P_{16}\big(M_{e}... M_{f}\otimes M_{g}... M_{h}\big)\in\TUCHM{16},\label{LU-2UCHMP16}
\end{equation}
where $...$'s are placeholders for two Fourier matrices $F_4(\alpha_1)$ and $F_4(\alpha_2)$ and two $\mathbb{I}_4$.
This time we obtain six two-parametric families each depending on two parameters $(\alpha_1,\alpha_2)\in[0,1)^{\times 2}$.

Since the orthogonal matrix $O_{16}$ (defined in Eq.~\eqref{O16_matrix}) is not a permutation matrix, the situation is slightly different.
There are only four one-parametric families, each depending on a single parameter $\alpha$, 
\begin{equation}
\big(M_{a}... M_{b}\otimes M_{c}... M_{d}\big)O_{16}\big(M_{e}... M_{f}\otimes M_{g}... M_{h}\big)\in\TUCHM{16},\label{LU-2UCHMO16}
\end{equation}
where $...$'s are placeholders for the only one Fourier matrix $F_4(\alpha)$ and three $\mathbb{I}_4$.
In particular, the ``simplest'' possible $2$-unitary real Hadamard matrix reads,
\begin{equation}
H=(F_2\otimes F_2 \otimes \mathbb{I}_4)O_{16},
\end{equation}
where $F_2$ denotes the standard Hadamard matrix $H_2$.

Consider a special dimension $d=4k$ with $k \in\mathbb{N}\setminus\{0\}$. 
In all such cases, AME$(4, d^2)$ states are represented by permutation matrices $P_{d^2}$~\cite{CD01}.
Assuming the Hadamard conjecture is true~\cite{Pa33, HW78}, we can also pick a real Hadamard matrix $H_d$.
We have
\begin{proposition}
Let $H_d$ be a real Hadamard matrix of order $d=4k$ for $k\geqslant 1$. Then there exist a $2$-unitary Hadamard matrix,\label{tensor-construction}
\begin{equation}
H_{d^2} = (H_d\otimes H_d)P_{d^2}\in\TUCHM{d^2}.\label{HddPdd}
\end{equation}
\end{proposition}
\label{prop:CHM_tensor}
That is, the tensor square of a real Hadamard matrix acting on a permutational representation of AME$(4,d^2)$ state brings it to the $2$-unitary form.
The veracity of this observation is an immediate consequence of the construction.
Moreover, this can be generalized to the complex domain by introducing monomial matrices as shown in Eqs.~\eqref{LU-2UCHM9} and~\eqref{LU-2UCHMP16},
where real Hadamard matrices $H_d$ can be replaced by appropriate complex counterparts. 

Note that for $d=6$, neither Prop.~\ref{tensor-construction}, nor
quantum cat map~\eqref{cat-map}, can provide any example of a $2$-unitary matrix in $\CHM{36}$.
During the preparation of the paper the problem of existence a two-unitary (complex) Hadamard matrix of order $36$ remained unsolved, however, recently, there appeared two reports containing an affirmative solution to this issue~\cite{Ra23, BZ24}.


\section{Multi-unitary Matrices and Absolutely Maximally Entangled States}
\label{sec:mu_AME}

This section contains information about multi-unitary matrices
in relation to quantum entanglement and a special class of quantum states. Also, we provide a physical motivation to investigate Hadamard matrices with a special internal structure.

Quantum entanglement is a property of Nature that describes strong correlations between physical systems at the sub-atomic level.
A primer and fundamental knowledge about this very special resource can be found in~\cite{HHHH09} and references therein.
By definition, entanglement needs at least two interacting systems to emerge.
Of special interest are interactions inside many-body systems which provide practical tools for the incoming era of quantum computing ---
quantum protocols, error correcting codes, and quantum cryptography require quantum entanglement.

Mathematical description of entanglement involves Hilbert space formalism.
In the commonly accepted interpretation, a composed quantum system
is modeled by a tensor product $\mathcal{H}=\mathcal{H}_A\otimes\mathcal{H}_B\otimes ...$, where
each $\mathcal{H}_X$ corresponds to individual subspace for each party.
Even in the simplest case of a
pure state $|\Psi\rangle$,
the question about its factorization onto the smaller components is a complex
and open problem.
In other words, if $|\Psi\rangle\neq |\psi_A\rangle\otimes |\psi_B\rangle\otimes ...$,
where $|\psi_X\rangle\in\mathcal{H}_X$, we call it an {\sl entangled state}, otherwise it is
{\sl separable}. 
Note that we restrict our attention to pure states only~\cite{BFZ21}. 
For mixed quantum states, the mechanism is similar~\cite{HHH96}, but beyond the scope of this paper.

In a general scenario, multipartite entanglement for a physical system
is described by the pair $(M,d)$ with $M$ being a number of parties (subsystems) and $d$ - number
of levels that each system admits.
We assume an equal number of degrees of freedom for each system, $\mathcal{H}_A=\mathcal{H}_B=...\equiv\mathcal{H}_d$ for $d\geqslant 2$, as well as an even number of parties $M=2k$.
A pure state $|\Psi\rangle$ of such a system is called 
{\sl absolutely maximally entangled} (AME) if it maximizes the entanglement among all equal bipartitions, that is tracing out $k=M/2$ subsystems should leave the state $\rho_{k}={\rm Tr}_{k}\rho_M$ in the {\sl maximally mixed} form, $\rho_{k}=\mathbb{I}_{d^{k}}/d^{k}$, where $\rho_M=|\Psi\rangle\langle\Psi |$
is a density matrix on the entire system.

In an alternative approach, a pure state $|\Psi\rangle$ is expressed by means of tensor of coefficients. Suppose $|\Psi\rangle$ is expanded in the product basis of $\mathcal{H}=\bigotimes \mathcal{H}_d$ as
\begin{equation}\label{eq:tensor_expansion}
    |\Psi\rangle = \sum_{i_1}^d\sum_{i_2}^d\cdots\sum_{i_M}^d\mathcal{T}_{i_1 i_2 ... i_M}\bigotimes_{j=1}^M|i_j\rangle,
\end{equation}
where $|i_j\rangle$ denotes $i^{\rm th}$ element of the standard basis of the $j^{\rm th}$ Hilbert space~\cite{BFZ21}.
A tensor $\mathcal{T}$ with $M$ indices can be reshaped into a matrix $U\in\mathbb{U}(d^k)$
by appropriate combinations of multi-index $\mathcal{J}=(j_1j_2...j_M)$. In general, there are $\frac{1}{2}\binom{M}{2}$ such rearrangements excluding global transpositions. If the matrix $U$ remains unitary after every possible rearrangement of $\mathcal{J}$, it is called {\sl $k$-unitary}. 
In particular, if the system contains $M=4$ subsystems divided pairwise onto two parts, $k=M/2=2$,
we recover well known notion of $2$-unitarity described in Sec.~\ref{sec:CHM_subsets}.
The concept of tensors with those properties, called {\sl perfect tensors},
was introduced in the context of solvable models for the bulk/boundary correspondence~\cite{PYHP15}. 
An AME state corresponding to a $k$-unitary permutation matrix $P_{d^{k}}$ is called a state with minimal support. 

A short list of multi-unitary matrices representing AME states of particular configuration
is presented below.

\begin{itemize}
\item There are no AME$(4,2)$ states~\cite{HS00}, which means that there are no 2-unitary matrices of size 4.
\item  A $3$-unitary Butson matrix $H_8$ corresponds to AME(6,2) state of six qubits~\cite{GALRZ15},
\begin{equation}
H_8={\scriptsize\left[\begin{array}{rr|rr|rr|rr}
                -&	-  &-   &+  &-  &+  &+  &+\\
                -&	-  &-	&+  &+  &-	&-	&-\\
                \hline
                -&	-  &+	&-  &-	&+	&-	&-\\
                +&	+  &-	&+	&-	&+	&-	&-\\
                \hline
                -&	+  &-	&-	&-  &-  &+  &-\\
                +&	-  &+	&+  &-	&-	&+  &-\\
                \hline
                +&	-  &-   &-	&+	&+	&+  &-\\
                +&	-  &-	&-  &-	&-	&-	&+\\
\end{array}\right]}\in\KUBH{8}{2}{3}\Longleftrightarrow {\rm AME}(6,2).
\end{equation}

\item Let $p$ be a prime number. There exist $3$-unitary matrices
$B\in\KUBH{p^3}{p}{3}$. In particular, there exists a Butson-type matrix 
corresponding to ${\rm AME}(6,3)$ state of six qutrits~\cite{He13}.

\item A $3$-unitary matrix matrix $H_{64}=P_{64}H_4^{\otimes 3}\in\KUCHM{64}{3}$ implies existence of the state ${\rm AME}(6,4)$ of six ququarts~\cite{He13}. The matrix $P_{64}$ is an appropriate $64$-dimensional permutation matrix.

\item There are no $4$-unitary matrices of order $16$ what follows from non-existence of the state AME$(8,2)$~\cite{Sc04}. 
\end{itemize}

A collection of AME states indicating their (non-)existence and additional brief commentary
is presented in the online service
maintained by Huber and Wyderka~\cite{HW23}.

\medskip

\subsection{\texorpdfstring{Nonexistence of Strongly $2$-Unitary Matrices}{}}

Finally, we complete the classification by considering the strong $2$-unitarity.
Matrix $X$ is said to be {\sl strongly $2$-unitary}
(or {\sl self $2$-unitary}), if $X=X^{\rm\Gamma}=X^{\rm R}$.
This can be generalized over strong $k$-unitary matrices which remain pairwise equal to unitary matrices
after any possible rearrangement of the multi-index (cf. previous section).

\begin{proposition}
There is no strongly $2$-unitary matrix $X_{d^2}$ of order $d^2$ for $d\geqslant 2$.
\label{no_self_2_unitary}
\end{proposition}
\begin{proof}
The case of $d=2$ is excluded by the Theorem {\bf 1}. in Ref.~\cite{HS00}.
In the general case, second column in $X_{d^2}=X_{d_2}^{\rm R}=X_{d^2}^{\rm\Gamma}$
is, by construction, repeated on the $(d+1)^{\rm th}$ position. Hence, matrix $X_{d^2}$ cannot have the full rank and be unitary.
\end{proof}
In particular, this means that strongly $2$-unitary complex Hadamard matrices do not exist either.


\section{Concluding Remarks}
Complex Hadamard matrices are interesting from the algebraic and combinatorial point of view. They are also often used in quantum information theory as they are building blocks of various 
quantum algorithms. 
Complex Hadamard matrices of size $d^2$ generate four-partite quantum states with a peculiar property: they are formed by a superposition of $d^4$ product states with equal weights, and from this perspective they are as quantum as possible.
Such matrices can be applied in several fields of theoretical physics including bulk/boundary correspondence~\cite{PYHP15}, quantum error correcting codes~\cite{Sc04} and quantum secret sharing~\cite{He13}.

In this paper we explored the
set of (complex) Hadamard matrices with a special structure, related to
absolutely maximally entangled states and multiunitary matrices.
The set of Hadamard matrices is rich enough and contains matrices
being (self) dual (${\rm R}$-dual),
${\rm\Gamma}$-dual and two-unitary.
This allows us to present several constructive analytic examples and even continuous multiparametric families of such objects. 
One exceptional case is strong $2^k$-unitarity, which is excluded just by construction -- in the simplest case, for $2$-unitarity, ($k=1$), there is no such matrix $X$ for which $X=X^{\rm R}=X^{\rm\Gamma}$.
Concerning $k$-unitarity, one of the open problems is the characterization of linear matrix transformations which do not spoil this property, as this would be of interest for quantum multipartite entanglement.

\section{Acknowledgements}
We acknowledge several fruitful discussions with Dardo Goyeneche and Suhail Ahmad Rather. We also wish to thank the participants of the \texttt{Hadamard 2022} conference~\cite{Hadamard2020} which took place at the Jagiellonian University in Kraków on the turn of June and July 2022, during which the problems presented in this paper were first raised. We are grateful to anonymous referees for their comments, especially for pointing out Ref.~\cite{GKL15}. WB is supported by NCN SONATA BIS grant no. 2019/34/E/ST2/00369. KŻ is supported by NCN Quantera project no. 2021/03/Y/ST2/0019.

\appendix

\section{Numerical techniques}\label{app:numerical_methods}

In pursuit of numerical examples, we used two different methods to obtain either a matrix $Y$ or a pair of two unitary diagonal matrices which bring a certain matrix $X$ into the form $Y=D_LXD_R$, with desired properties.

Let $D_L$ and $D_R$ be two unitary diagonal matrices, each depending on $N$ real phases
$\alpha=(\alpha_1,...,\alpha_N)$ and 
$\beta=(\beta_1,...,\beta_N)$, respectively.
The first method is a random walk over the phases $\alpha$ and $\beta$. Define a function $\mathcal{Z}$ depending on $X$ and $2N$ real variables:
\begin{equation}
\mathcal{Z}(X,\alpha_1,...,\alpha_N;\beta_1,...,\beta_N) = \chi\big(D_L(\alpha)XD_R(\beta)\big),\label{Z_function}
\end{equation}
where
\begin{equation}
\chi(U)=|S(U)-1|+|S(U^{\rm\Gamma})-1|+|S(U^{\rm R})-1| 
\end{equation}
with $S$ being a linear entropy defined in Eq.~\eqref{SL_entropy} for any matrix, in particular for $U\in\mathbb{U}(N)$, while $X$ in Eq.~\eqref{Z_function} represents some element from $\CHM{N}$ for $N=9$ or $N=16$.
For many $X\in\CHM{N}$ or $X\in\BH{N}{q}$ the minimizing procedure
quickly converges, which is equivalent to $\mathcal{Z}\to 0$. 
Possibility to fix some phases allows the recovery of
analytical forms of $D_L$ and $D_R$. It is also easy to modify function $\chi(U)$ in order to search for dual or self-dual CHM.

When one needs to obtain a single matrix from
the set $\TUCHM{N}$ or $\DUCHM{N}$, another numerical recipe should be applied.
Recall the Sinkhorn algorithm, which was originally designed to generate bistochastic matrices~\cite{Si64}.
Convergence of this alternating combination of disjoint operations
is assured under strict mathematical conditions~\cite{BB96}, however, the flexibility of this method
proved to be useful in a wide range of applications.
Recently the modified version of this procedure was successfully used in search of the very special
quantum states describing multipartite entanglement~\cite{RBBMLZ22} and it assisted
the discovery of a series of new complex Hadamard matrices~\cite{Br23} in several dimensions $N\geqslant 8$.
Here we present yet another modification adapted for $2$-unitary CHM.

Consider a map $\mathcal{M}_S:\mathbb{C}^{N^2}\to\CHM{N}$ defined by an iterative procedure.
We start with a matrix $X_0$ (a seed) with entries from the random Gaussian distribution,
$\forall\,j,k:(X_0)_{jk} \in \mathcal{N}(0,1)$, and every next iteration consists of four operations: 1) normalization of entries, 2)
polar decomposition (${\rm P}_{\rm d}$), 3) reshuffling ($^{\rm R}$),
and 4) partial transpose ($^{\rm\Gamma}$). This can be written compactly with help of an intermediate matrix $T_n$ as
\begin{equation}
\mathcal{M}_S:\begin{cases}
        (T_n)_{jk} &= \dfrac{(X_n)_{jk}}{|(X_n)_{jk}|},\\
        \\
        X_{n+1} &=  \big({\rm P}_{\rm d}(T_n)\big)^{\rm \Gamma R} \quad \text{for} \quad n\geqslant 1.
\end{cases}
\end{equation}
In many cases, as $n\to\infty$, for a random seed $X_0$, we obtain $Y=\mathcal{M}_S(X_0)$ being a $2$-unitary CHM of dimension $N=9$ or $N=16$.


\section{\texorpdfstring{Examples of Self-Dual (R-Dual) and Self $\Gamma$-Dual CHM}{}}\label{app:self_dual}
A four-parametric self-${\rm R}$-dual unitary family stemming from the Fourier matrix $F_9^{(4)}$~\cite{TZ06}
can be obtained by using two diagonal matrices,
\begin{align}
    D_L &= {\rm diag}\Big\{1, 1, 1, 1, 1, 1, e^{2i\pi \alpha}, e^{2i\pi \alpha}, e^{2i\pi \alpha}\Big\} : \alpha \in [0, 1),\\
    D_R &= {\rm diag}\Big\{1, 1, 1, 1, a\omega^2, b\omega^4, 1, c\omega^4, d\omega^8\Big\} : \omega = \exp\{i\pi/9\},
\end{align}
as $Y=D_L(\alpha) F_9^{(4)}(a,b,c,d) D_R\in\DUCHM{9}$.
The matrix $D_R$ depends on the original parameters $\{a,b,c,d\}$ involved into the Fourier one,
while $D_L$ depends on an additional parameter which does not affect self-duality. 
Entropy $S(Y)=(1,1,f(a,b,c,d))$, where $f$ is a function of independent parameters.

With appropriate diagonal unitary matrices $D_L$ and $D_R$, it is possible to recover numerous other similar examples, including $D_L(F_3\otimes F_3)D_R$,
 $D_L(F_4(a)\otimes F_4(a))D_R$, $D_LH_{16A,B}D_R$, $D_L(F_2\otimes F_8^{(5)}(a,b,c,d,e))D_R$, which belong to $\DUCHM{9}$ or $\DUCHM{16}$.

Similarly, if the reshuffling operation is replaced by partial transpose, one
can discover the following matrices.

Isolated Butson matrix of order $N=9$,
\begin{equation}
B_9^{(0)}=\exp\Bigg\{\frac{i\pi}{3}{\scriptsize\left[\begin{array}{ccc|ccc|ccc}
. & . & . & . & . & . & . & . & .\\
. & 5 & 3 & 2 & 5 & 3 & 2 & 1 & 5\\
. & 3 & 3 & . & 1 & 5 & 4 & 1 & 3\\
\hline
. & 2 & . & 2 & . & 2 & 4 & 4 & 4\\
. & 5 & 1 & . & 3 & 3 & 4 & 3 & 1\\
. & 3 & 5 & 2 & 3 & 5 & 2 & 5 & 1\\
\hline
. & . & 2 & 4 & 2 & . & 2 & 4 & 4\\
. & 3 & 3 & 4 & 5 & 1 & . & 3 & 1\\
. & 1 & 5 & 4 & 3 & 3 & . & 1 & 3
\end{array}\right]}\Bigg\}\in\BH{9}{6}
\end{equation}
becomes self $\Gamma$-dual when transformed by two diagonal unitary matrices of the form
$D_L={\rm diag}\{1,1,1,1,\omega,1,1,1,\omega\}$ and 
$D_R={\rm diag}\{1,1,1,1,\omega,1,1,\omega,\omega^2\}$
with $\omega=\exp\{2 i\pi/3\}$. That is
$Y=D_LB_9^{(0)}D_R$, $S(Y)=(1,\frac{20}{27},1)$ and $Y=Y^{\rm\Gamma}\in\GUBH{9}{6}$. Note that in this class of matrices, the entropy ceases to be flat on the central component.

Another isolated example can be constructed out of the well-known matrix $N_9^{(0)}$ found by Beauchamp and Nicoară~\cite{BN06}. Here, we
present its equivalent form
\begin{equation}
N_9^{(0)}\simeq{\scriptsize\left[\begin{array}{rrrrrrrrr}
                1&	1     &1	&1      &1	    &1	    &1	    &1  	&1\\
                1&	y     &y^2	&-1     &-y	    &y   	&y^3	&y^3	&y\\
                1&	y^2   &y^4	&-y     &y^2	&-y^3	&y^4	&y^2	&1\\
                1&	-y^4  &-y^3	&y^3	&-y^3	&-y^2	&-y^4	&-y^2	&-1\\
                1&	-1    &y^2	&-1/y	&1	    &1	    &y	    &y^2	&1/y\\
                1&	y^3   &-y	&-1     &y^2	&y^3	&y	    &y	    &y\\
                1&	y     &1	&-1/y	&y^2	&y^2	&-1	    &1	    &1/y\\
                1&	y^4   &y^2	&-y     &y^4	&y^2	&y^2	&-y^3	&1\\
                1&	y^3   &y^4	&-y^3	&y^4	&y^2	&y^3	&y^2	&-1
\end{array}\right]}, \,\,\,\, y = -\frac{1}{4} + i\frac{\sqrt{15}}{4}.
\end{equation}
Two diagonal matrices
\begin{align}
D_L&={\rm diag}\big\{1,1,1,1,-y^4,-y^3,1,y,1\big\},\\
D_R&={\rm diag}\big\{1,1,1,1,-1,-y,-y^3,\xi,\xi y\big\} \quad : \quad \xi = \frac{7}{2^7}+i\frac{33\sqrt{15}}{2^7}
\end{align}
bring $N_9^{(0)}$ to $Y=D_LN_9^{(0)}D_R$ so that
$S(Y)=(1,\frac{22259}{31104},1)$
and $Y=Y^{\rm\Gamma}\in\GUCHM{9}$. The fact
we need to introduce a rather uncommon unimodular number $\xi$ remains the mystery of complex Hadamard matrices.

Not only isolated matrices provide self $\Gamma$-dual structures in $\CHM{9}$. The Karlsson's matrix $K_9^{(2)}(\zeta)$ with $\zeta=3$, 
$D_L=D_R={\rm diag}\{1,1,1,1,\omega^2,\omega,\omega,\omega^2,1\}$
and $\omega=\exp\{2i\pi/3\}$ defines a $\rm\Gamma$-dual matrix $Y=D_LK_9^{(2)}(3)D_L$ such that $S(Y)=(1,\frac{3}{4},1)$
and $Y=Y^{\rm\Gamma}\in\GUCHM{9}$.
The tensor product of two Fourier matrices $F_3\otimes F_3$ can be turned
into a self $\Gamma$-dual matrix,
\begin{equation}
Y(\alpha)=\left({\rm diag}\big\{1,1,e^{2i\pi\alpha}\big\}\otimes \mathbb{I}_3\right)\left(F_3\otimes F_3\right)
\end{equation}
to form a one-parametric family
such that $Y(\alpha)=Y_9^{\rm\Gamma}(\alpha)$ and $S(Y(\alpha))=(1,0,1)$ for $\alpha\in[0,1)$.
Similarly, one has $Y(a_1,a_2)=F_4(a_1)\otimes F_4(a_2)\in\GUCHM{16}$ without any additional tuning by diagonal matrices;
$Y(a_1,a_2)=Y^{\rm\Gamma}(a_1,a_2)$ and $S(Y(a_1,a_2))=(1,0,1)$ for all $a_1,a_2\in[0,1)$. 

Finally, the following list contains self $\Gamma$-dual CHM obtained by the tensor product construction of two Fourier matrices $F_N$ for $5\leqslant N\leqslant 16$ including possible parameterizations:
\begin{align}
Y_{25}&=F_5^{(0)}\otimes F_5^{(0)},\\
Y_{36}(\alpha_1,\alpha_2)&=F_6^{(2)}(\alpha_1,\alpha_1)\otimes F_6^{(2)}(0,0),\\
Y_{49}&=F_7^{(0)}\otimes F_7^{(0)},\\
Y_{64}(\alpha_1,...,\alpha_5)&=F_8^{(5)}(\alpha_1,...,\alpha_5)\otimes F_8^{(5)}(0,0,0,0,0),\\
Y_{81}(\alpha_1,...,\alpha_4;\alpha_5,\alpha_6)&=F_9^{(4)}(\alpha_1,...,\alpha_4)\otimes F_9^{(4)}(\alpha_5,0,0,\alpha_6),\\
Y_{100}(\alpha_1,...,\alpha_4)&=F_{10}^{(4)}(\alpha_1,...,\alpha_4)\otimes F_{10}^{(4)}(0,0,0,0),\\
Y_{121}&=F_{11}^{(0)}\otimes F_{11}^{(0)},\\
Y_{144A}(\alpha_1,...,\alpha_9;\alpha_{10})&=F_{12A}^{(9)}(\alpha_1,...,\alpha_9)\otimes F_{12A}^{(9)}(\alpha_{10},0,0,0,0,0,0,0,0),\\
Y_{144B}(\alpha_1,...,\alpha_9)&=F_{12B}^{(9)}(\alpha_1,...,\alpha_9)\otimes F_{12B}^{(9)}(0,0,0,0,0,0,0,0,0),\\
Y_{144C}(\alpha_1,...,\alpha_9;\alpha_{10})&=F_{12C}^{(9)}(\alpha_1,...,\alpha_9)\otimes F_{12C}^{(9)}(\alpha_{10},0,0,0,0,0,0,0,0),\\
Y_{144D}(\alpha_1,...,\alpha_9)&=F_{12D}^{(9)}(\alpha_1,...,\alpha_9)\otimes F_{12D}^{(9)}(0,0,0,0,0,0,0,0,0),\\
Y_{169}&=F_{13}^{(0)}\otimes F_{13}^{(0)},\\
Y_{196}(\alpha_1,...,\alpha_6)&=F_{14}^{(6)}(\alpha_1,...,\alpha_6)\otimes F_{14}^{(6)}(0,0,0,0,0,0),\\
Y_{225}(\alpha_1,...,\alpha_8)&=F_{15}^{(8)}(\alpha_1,...,\alpha_8)\otimes F_{15}^{(8)}(0,0,0,0,0,0,0,0),\\
Y_{256}(\alpha_1,...,\alpha_{17})&=F_{16}^{(17)}(\alpha_1,...,\alpha_{17})\otimes F_{16}^{(17)}(0,...,0).
\end{align}
In each case, $\alpha_j\in[0,1)$, $S(Y_N)=(1,0,1)$ and $Y_N=Y_N^{\rm\Gamma}$. 
Moreover, in general, for each dimension $N$, matrix $Y_{N^2}:=F_N(\vec{0}) \otimes F_N(\vec{0})$ is self $\rm\Gamma$-dual, where $\vec{0}$ represents the vector of parameters $\alpha_j$ set to 0 for all $j$, provided a given Fourier matrix admits additional parameterization.


\section{\texorpdfstring{Family of $2$-unitary CHM of order $N=16$}{}}\label{app:2UCHM_16}

The Sinkhorn algorithm produces several numerical examples of $2$-unitary complex Hadamard matrices in the set $\TUCHM{16}$. Many of them admit free parameters and can be expressed analytically.
Here we present a single, one-parameter affine family in the form of its $15$-dimensional core,
\begin{align}
&{\rm core}\Big(T_{16}^{(1)}(a)\Big)=\\
&{\scriptsize\left[\begin{array}{rrr|rrrr|rrrr|rrrr}
-1&  -i a  &   a  &  -a   &  i a &  1  &  -1  &   1   & -1   & -a   &  i a  &   1  &  -1   &  a   & -i a  \\
 i&   a    &  -a  &   a   & -a   & -i  &  -1  &  -i   & -1   &  a   & -a    &   i  &   1   &  a   & -a    \\
 1&  -1    &  -1  &  -1   & -1   &  1  &   1  &   1   &  1   & -1   & -1    &   1  &   1   & -1   & -1    \\
 \hline
-1&   1    &  -1  &  -1   &  1   & -1  &   1  &   1/a &  i/a & -i a &  a    &  -1/a&  -i/a &  i a & -a    \\
 1&   i a  &   a  &  -a   & -i a & -1  &  -1  &  -i   &  i   & -i a & -a    &  -i  &   i   &  i a &  a    \\
-i&  -a    &  -a  &   a   &  a   &  i  &  -1  &  -1   &  i   &  i a & -i a  &   1  &  -i   &  i a & -i a  \\
-1&  -1    &   1  &   1   & -1   & -1  &   1  &  -1/a & -i/a & -i a &  a    &   1/a&   i/a &  i a & -a    \\
\hline
-i&  -i b  &  -b  &   i a & -i a &  i a&  -i a&   i   & -1   &  b   &  i b  &  -i a&   i a & -i a &  i a  \\
-1&   i a  &  -a  &  -i a &  a   & -i  &  -i  &  -1   &  1   & -a   &  i a  &   i  &   i   & -i a &  a    \\
 i&  -a    &   a  &   i a &  i a & -1  &  -i  &   i   &  1   &  a   & -a    &  -1  &  -i   & -i a & -i a  \\
-i&   i b  &   b  &   i a & -i a & -i a&   i a&   i   & -1   & -b   & -i b  &   i a&  -i a & -i a &  i a  \\
\hline
 1&   1    &   1  &  -i a & -a   &  1/a&  -i/a&  -1/a &  i/a &  i a &  a    &  -1  &  -1   & -1   & -1    \\
-1&  -i a  &   a  &  -i a &  a   &  i  &   i  &  -i   & -i   &  i a & -a    &  -1  &   1   & -a   &  i a  \\
 i&   a    &  -a  &   i a &  i a &  1  &   i  &  -1   & -i   & -i a & -i a  &  -i  &  -1   & -a   &  a    \\
 1&  -1    &  -1  &  -i a & -a   & -1/a&   i/a&   1/a & -i/a &  i a &  a    &  -1  &  -1   &  1   &  1    \\
\end{array}\right]},\nonumber
\end{align}
where $b=a^2$, while $a=\exp\{2i\pi p_1\}$ is a unimodular parameter depending on $p_1\in[0,1)$. 
The matrix $Y_{16}^{(1)}(a)\equiv D_L(a)T_{16}^{(1)}(a)D_R(a)$, with $\omega=\exp\{i\pi/6\}$ and 
\begin{align}
D_L(a) &= {\rm diag}\big\{1,1,1,1,1,1,-1,1,1,a,-a,-1,1,-1,-1,1\big\},\\
D_R(a) &= {\rm diag}\big\{1,1,1,1,1,-i,-ia^2,-a^2,\omega,\omega^4,\omega^7,\omega^{10},-i,-1,i,1\big\},
\end{align}
forms a $2$-unitary family in $\TUCHM{16}$.
Several other examples can be found on GitHub~\cite{GitHub}.


\end{document}